\documentstyle[preprint,aps]{revtex}  

\begin{document}
\thispagestyle{empty}
\draft         
\preprint{hep-ph/9306229, ~IFT-P.014/93 ~and~ IFUSP-P 1042}
\title{ Searching for Leptoquarks in electron-photon Collisions}
\author{ O.\ J.\ P.\ \'Eboli $^a$, E.\ M.\ Gregores $^b$, M.\ B.\
Magro $^a$, \\
P.\ G.\ Mercadante $^a$  and S.\ F.\ Novaes $^b$}
\address{$^a$ Instituto de F\'{\i}sica, Universidade de S\~ao Paulo \\
C.P. 20516, 01498-970 S\~ao Paulo, Brazil \\
$^b$ Instituto de F\'{\i}sica Te\'orica, Universidade Estadual 
Paulista\\
Rua Pamplona 145, 01405-900 S\~ao Paulo, Brazil}
\date{\today}
\maketitle
\begin{abstract}
We study the production of composite scalar leptoquarks in $e\gamma$
colliders, and we show that an $e^+e^-$ machine operating in its
$e\gamma$ mode is the best way to look for these particles in $e^+e^-$
collisions, due to the hadronic content of the photon. 
\vskip 2.5cm
\begin{center}
{\em Submitted to Phys.\ Lett.\ B}
\end{center}
\end{abstract}

\newpage

The reason for the observed pattern of the fermionic generations is not
elucidated in the framework of the $SU(3)_C \otimes SU(2)_L \otimes U(1)_Y$
Model. Nevertheless, there are theories, beyond the Standard Model that suggest
a natural explanation for the existence of lepton and quark families, invoking
a deeper structure which is common to both kind of fermions. Some composite
models \cite{af}, for instance, suggest a preonic sub-structure, where quarks
and leptons have some common constituents. This class of models exhibits a very
rich spectrum which includes excited states of the known particles as well as
new particles possessing unusual quantum numbers. Some of these models predict
the existence of leptoquarks, which are color triplets and carry simultaneously
leptonic and baryonic number. These particles are also present in several other
theories beyond the Standard Model such as technicolor models \cite{tec}, grand
unified theories \cite{gut}, and superstring-inspired models \cite{e6}.

The production and signals of leptoquarks have already been analyzed in the
literature for $ep$ \cite{wud,lepto:ep,buch}, hadronic \cite{lepto:ha}, and $e^+e^-$
colliders \cite{lepto:ee1,lepto:ee2}.  The study of single leptoquark
production in $e\gamma$ collisions was carried out in Ref.\ \cite{lepto:ee2}
without considering the resolved photon contribution. In this work, we study
the capability of an $e\gamma$ collider to search for these particles, taking
into account the hadronic content of the photon. We demonstrate that the next
generation of linear $e^+e^-$ colliders, operating in the $e\gamma$ mode, is
the best place to look for these particles in such colliders. An $e\gamma$
machine, due to the quark and gluon distributions of the photon, is able to
exhibit a very rich initial state which allows a detailed study of these
particles. 

We shall consider an $e\gamma$ collider where the hard photons are
obtained by the laser backscattering mechanism, which converts an $e$
beam into a $\gamma$ one \cite{las0,laser}.  The basic idea is to
scatter soft photons of a few eV laser from one of the energetic beams
of an  $e^+e^-$ linear collider. This mechanism is able to produce a
collimated and energetic photon beam which has nearly the same
luminosity and energy of the parent electron beam.

The production of leptoquarks in $e\gamma$ collisions can occur either
by direct or ``resolved'' photon processes, i.e. with the photon
interacting  through its hadronic content. In the latter case it is
possible, for instance, to produce resonant leptoquarks through the
quark component of the photon. We shall study the following signals
for the leptoquark $S$
\begin{equation} 
e + q_\gamma  \rightarrow  e + \text{jet} \; ,
\label{e:jet}
\end{equation}
\begin{equation} 
e + q_\gamma \rightarrow  \mbox{jet} + \mbox{E}_{\text{missing}} \; ,
\label{nu:jet}
\end{equation}
\begin{equation} 
e + \gamma (g_\gamma)  \rightarrow  S + \mbox{jet} \; ,
\label{s:jet}
\end{equation}
where $q_\gamma$ and $g_\gamma$ denotes the quark and gluon content of
the photon respectively. The processes (\ref{e:jet}) and
(\ref{nu:jet}) contain the production of a leptoquark in the $s$
channel, while the process (\ref{s:jet}) is the associate production
of a leptoquark and a quark jet.

For the sake of definiteness, we shall consider scalar leptoquarks predicted by
the Abbott--Farhi model \cite{af},   which is a confining version of the
Standard Model and it is also called Strongly Coupled Standard Model (SCSM). 
The SCSM is described by a Lagrangian which has the same general structure of
the Standard Model one.  Nevertheless, no spontaneous symmetry breaking occurs
and the $SU(2)_L$ gauge interaction is confining.  In this model, the physical
left-handed fermions are bound states of a preonic scalar and a preonic
dynamical left-handed fermion, while the vector bosons are P--wave bound states
of the scalar preons. 

The SCSM model cannot be analyzed perturbatively since it is strongly
interacting in the energy scale of interest. Therefore, we shall parametrize the
interaction between leptoquarks and physical left-handed fermions by an
effective Lagrangian \cite{wud} given by
\begin{equation}
{\cal L} = \frac{\lambda^{ab}}{2} 
\left( S^\dagger_{ab}~  L_a^T~  C~ \tau_2~ L_b~ +~  \text{h.c.} \right) \; ,
\label{l:wud}
\end{equation}
where $S_{ab}$ are the scalar leptoquark fields, $L_a$ are the
physical left-handed lepton and quark SU(2) doublets ($a,b = 1,
\cdots, 12$), $C$ is the charge conjugation matrix, and $\lambda^{ab}$
are dimensionless coupling constants.  Assuming that the leptoquark
couples only with fermions of a given generation, in $e\gamma$
collisions, we can restrict ourselves to the sector of the Lagrangian
that describes the leptoquark couplings to $(\nu_e, e)_L$ and $(u,
d)_L$, i.e.
\begin{equation} 
{\cal L} = \lambda~ i~
S^\dagger \left(
\bar{e}^c_L u_L -  \bar{\nu}^c_L  d_L \right) + \text{h.c.} \; .
\label{l:first}
\end{equation}
This Lagrangian is a prototype of a wide class of models \cite{buch}
presenting the interaction of color triplet, charged scalars with $B$
and $L$ numbers, which is diagonal in the generation. There are few
constraints from low energy phenomenology on models when leptoquarks
are generational diagonal and the couplings to fermions are chiral
\cite{buch:2}.  Recent experimental results from CDF \cite{cdf} were
able to establish a lower bound of $M_S > 116$ GeV, assuming the
branching ratio $BR(S\rightarrow eq) = 1$.

The total cross section for leptoquark production in an
$e\gamma$ collider can be obtained by folding the elementary
cross section for the processes leading to the signals 
(\ref{e:jet}--\ref{s:jet}) with the electron-parton
luminosity.   
\begin{equation}
\sigma_{e^+ e^- \rightarrow e \gamma(p) \rightarrow X} (s)=
\int_{z_{\text{min}}}^{1} dz ~ \frac{d{\cal L}_{ep}}{dz} 
\hat\sigma_{e  \gamma(p) \rightarrow X} (\hat s) \; ,
\label{sig:tot}
\end{equation}
where $z^2= \tau = \hat{s}/s$, with $s$ ($\hat{s}$) being the
center-of-mass energy squared  of the $e^+e^-$ (electron-parton)
system, and $d{\cal L}_{ep}/dz$ stands for the electron-parton
($p=\gamma, q, g$) differential luminosity.

For direct photon processes, the $e\gamma$ luminosity is
\begin{equation}
\frac{d {\cal L}_{e\gamma}}{dz} = 2 z F_{\gamma/e} (z^2) \; ,
\label{l:eg}
\end{equation}
where the energy
spectrum of backscattered laser photons ($F_{\gamma/e}$)
is \cite{laser}   
\begin{equation} 
F_{\gamma/e} (x)  = 
\frac{1}{D(\xi)} \left[ 1 - x + \frac{1}{1-x} -
\frac{4x}{\xi (1-x)} +   \frac{4 x^2}{\xi^2 (1-x)^2}  \right] \; , 
\label{F:laser}
\end{equation}
with $x$ being the fraction of the electron energy carried by
the photon, and
\begin{equation}
D(\xi) = \left(1 - \frac{4}{\xi} - \frac{8}{\xi^2}  \right) 
\ln (1 + \xi) + \frac{1}{2} + \frac{8}{\xi} - \frac{1}{2(1 + \xi)^2}
\; , 
\label{D}
\end{equation}
with $\xi  \simeq 2 \sqrt{s} \omega_0/m^2$,  where $\omega_0$ is the
laser photon  energy,  and $m$ the electron mass. In (\ref{l:eg}) we
have assumed that the average number of high energy converted photons
per electron is equal to one. The fraction of photons with energy
close to the maximum value grows with $E$ and $\omega_0$.
Nevertheless, $\omega_0$ is constrained since the laser energy should
not exceed the threshold for $e^+e^-$ pair creation via the
interaction of the backscattered and laser photons, otherwise the
conversion $e \rightarrow \gamma$ would be reduced. We have taken into
account this constraint in our numerical evaluations.

Interactions initiated by ``resolved" photons are described in terms of
structure function of partons (quarks and gluons) inside the photon
\cite{witten}. We can define the electron-parton luminosity by folding the
photon structure functions with the photon distribution in the electrons,
\begin{equation} 
\frac{d {\cal L}_{ep}}{dz} = 2 z \int_{z^2}^1 \frac{dx}{x}
F_{\gamma/e} (x) P_\gamma (z^2/x,Q^2) \; ,  
\label{par:pho}
\end{equation} 
where $P_\gamma = Q_\gamma$ $(G_\gamma)$ is the quark (gluon)
structure function. There are several parametrizations for these
structure functions available in the literature \cite{do,dg,lac}. We
present our results for the fitting of Drees-Grassie (DG) \cite{dg}
and we checked that they do not change in a significant way if we use,
for instance, the set from Levy-Abramowicz-Charchula (LAC3)
\cite{lac}, which is  characterized by a harder gluon spectrum. We
evaluated all the structure functions and the strong coupling constant
($\alpha_s$) at a scale $Q^2 = \hat s/4$. 

Let us start our analyses by the signal of an electron accompanied by
a jet. Taking into account the couplings of Eq.(\ref{l:first}), this
signal is associated to the subprocess $eq(\bar q) \rightarrow eq(\bar
q)$ with $q = u$, $d$.  In the case of $eu\rightarrow eu$, there are
contributions from the exchange of the leptoquark $S$ in the
s-channel, and from $\gamma$ and $Z$ in the t-channel \cite{cor},
leading to
\begin{eqnarray} 
\left(\frac{d\hat{\sigma}}{d\hat{t}}\right)_{eu \rightarrow eu} &=&
\frac{1}{16\pi} \left\{ \left( \frac{-2e^{2}}{3\hat{t}} + 
\frac{R_e R_u }{ \hat{t}-M_{Z}^{2}}\right) ^{2} 
+ \left( \frac{\lambda^{2}}{2} \frac{M_S\Gamma_S}
{ \left( \hat{s}-M_S^{2}\right) ^{2}+M_S^{2}\Gamma_S^{2}}\right)^{2} 
\right.  \nonumber \\ 
& & + \left( \frac{-2e^{2}}{3\hat{t}} + 
\frac{L_e L_u}{\hat{t}-M_{Z}^{2}} - \frac{\lambda^{2}}{2}
\frac{\hat{s}-M_S^{2}} { \left( \hat{s}-M_S^{2}\right)
^{2}+M_S^{2}\Gamma_S^{2}} \right) ^{2}  \nonumber \\ 
&&\left. + \frac{\hat{u}^{2}}{\hat{s}^2} \left[ 
\left(\frac{-2e^{2}}{3\hat{t}} +  \frac{ R_e L_u }
{\hat{t}-M_{Z}^{2}}\right)^{2} + \left( \frac{-2e^{2}}{3\hat{t}} +
\frac{L_e R_u}{\hat{t}-M_{Z}^{2}} \right) ^{2} \right]\right\} \; ,
\label{eu:eu}  
\end{eqnarray}
where,
\begin{equation}
L_f = \frac{e}{\sin\theta_W \cos\theta_W } 
\left( T^3_f - Q_f \sin^2\theta_W  \right) \; , \;\;\;
R_f = - e Q_f \tan\theta_W \; ,
\label{g:rl}
\end{equation}
with $Q_f$ and $T^3_f$ being the fermion charge in units of the proton
charge, and the third component of the weak-isospin, respectively.
The subprocess $e\bar{u} \rightarrow e\bar{u}$ receives contributions
from the exchange of $\gamma$ and $Z$ in the t-channel and $S$ in the
u-channel, resulting
\begin{eqnarray}
\left(\frac{d\hat{\sigma}}{d\hat{t}}\right)_{e\bar{u}\rightarrow 
e\bar{u}} = \frac{1}{16 \pi} \left\{
\left(\, \frac{-2e^{2}}{3\hat{t}} +  \frac{R_e
L_u}{\hat{t}-M_{Z}^{2}} \,\right) ^{2} + \left(\,
\frac{-2e^{2}}{3\hat{t}} +  \frac{L_e R_u}{\hat{t}-M_{Z}^{2}}
\,\right) ^{2}  \right.
 \nonumber \\
\left. + \frac{\hat{u}^{2}}{\hat{s}^2}\left[ \left(\, 
\frac{-2e^{2}}{3\hat{t}} + 
\frac{R_e R_u}{\hat{t}-M_{Z}^{2}} \,\right) ^{2}
+ \left(\, \frac{-2e^{2}}{3\hat{t}} +  \frac{L_e
L_u}{\hat{t}-M_{Z}^{2}} - \frac{\lambda^{2}}{2\left(
\hat{u}-M_S^{2}\right) } \,\right) ^{2} \right] \right\} \; .
\label{eubar:eubar}
\end{eqnarray}
The subprocesses $e d(\bar{d}) \rightarrow e d(\bar{d})$ does not
involve the leptoquark $S$, however it should be included in the
evaluation of the electron plus jet signal, since it is an irreducible
background for this process.
\begin{eqnarray}
\left(\frac{d\hat{\sigma}}{d\hat{t}}\right)_{ed \rightarrow ed} 
&=& \frac{1}{16 \pi} \left\{
\left(\, \frac{e^{2}}{3\hat{t}} + 
\frac{R_e R_d}{\hat{t}-M_{Z}^{2}}\,\right) ^{2} 
+ \left(\, \frac{e^{2}}{3\hat{t}} +  
\frac{L_e L_d}{\hat{t}-M_{Z}^{2}} \,\right) ^{2}  \right.
 \nonumber \\
&&\left. + \frac{\hat{u}^{2}}{\hat{s}^2}\left[ \left(\, 
\frac{e^{2}}{3\hat{t}} + 
\frac{R_e L_d}{\hat{t}-M_{Z}^{2}} \,\right) ^{2}
+ \left(\, \frac{e^{2}}{3\hat{t}} +  \frac{L_e
R_d}{\hat{t}-M_{Z}^{2}} \,\right) ^{2} \right] \right\}
\label{ed:ed}
\end{eqnarray}
The result for $e \bar{d} \rightarrow e \bar{d}$ can be
obtained from Eq. (\ref{ed:ed}) by the interchange $L_d
\leftrightarrow R_d$.

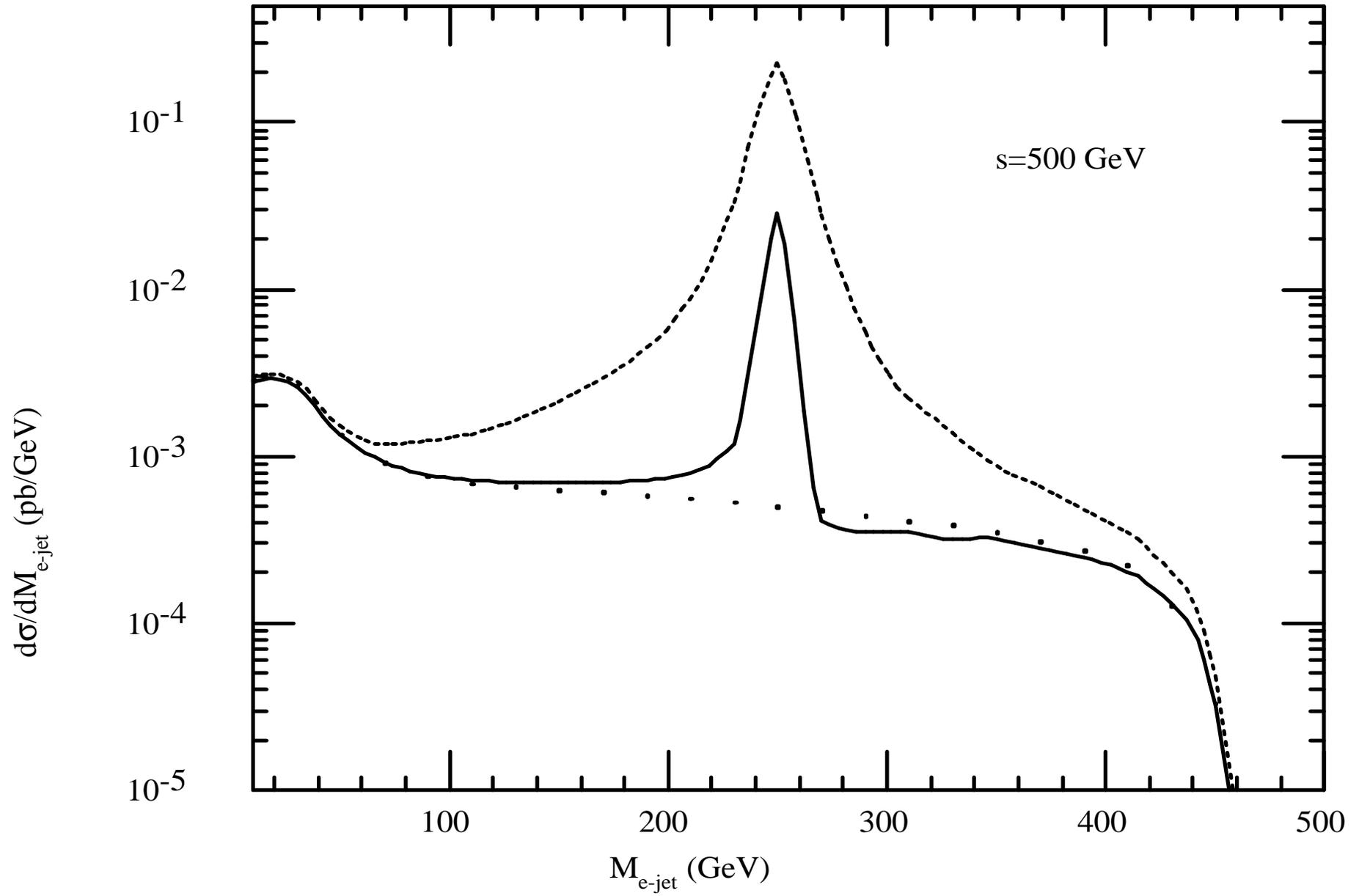
\begin{figure}
\protect
\caption{Invariant mass distribution for the process $e^+ e^-
\rightarrow e + q_\gamma \rightarrow e + \mbox{jet}$. We considered  an
$e^+ e^- $ collider with $\protect\sqrt{s} = 500$ GeV and fixed $M_S =
250$ GeV. The curves are for the leptoquark coupling $\lambda = e$
(solid line), and  $\lambda = 1.0$  (dashed line). The  dotted line
represents the Standard Model prediction i.e. $\lambda=0$.} 
\label{fig:1}
\end{figure}

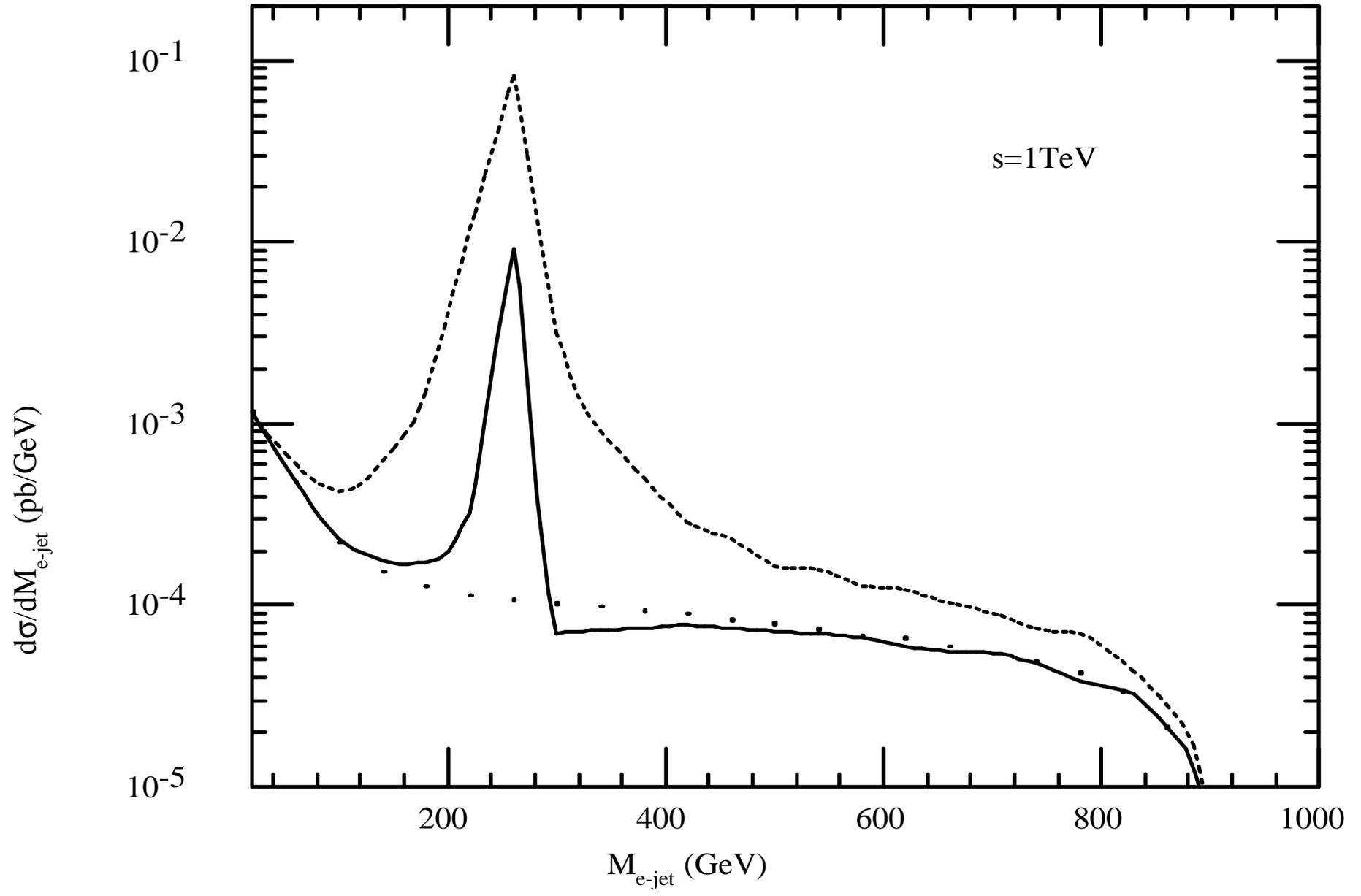
\begin{figure}
\protect
\caption{The same as Figure \protect\ref{fig:1} for  
$\protect\sqrt{s} = 1000$ GeV.}
\label{fig:2}
\end{figure}

In Figs. \ref{fig:1} and \ref{fig:2} we show the $e-$jet invariant
mass distribution for the process $e\gamma \rightarrow e +
\mbox{jet}$, for $\sqrt{s} = 500$ and $1000$ GeV respectively.  We
assumed $M_S = 250$ GeV and we plotted the result for different values
of $\lambda$, in particular, the choice $\lambda=0$ corresponds to the
Standard Model prediction.  Since the planned machines will have a
luminosity of the order of $10^4$ pb$^{-1}$ per year, we can learn
from these figures that it will be an easy task to discover the
leptoquark in this mode, since there is a very clear peak  well above
the Standard Model background. Moreover we will be able to obtain a
very precise measurement of $M_S$ and $\lambda$ due to the high
statistic of the planned machines.

The signal (\ref{nu:jet}) can be obtained through the subprocess $e u
\rightarrow \nu_e d$ with contributions of $W$ boson  in the t-channel
and of the leptoquark $S$ in the s-channel, yielding
\begin{eqnarray}
\left(\frac{d\hat{\sigma}}{d\hat{t}}\right)_{e u \rightarrow \nu_e d} 
&=& \frac{1}{64 \pi}
\left[ \left(\, \frac{ g^{2} }{ \hat{t}-M^{2}_{W} } +
\frac{\lambda^{2}\left(\hat{s}-M_S^{2}\right)}
{\left(\hat{s}-M_S^{2}\right)^{2}  + M_S^{2}\Gamma_S^{2} } \,\right)
^{2} \right.  
 \nonumber \\ 
&&+ \left. \left(\, \frac{ \lambda^{2}
M_S \Gamma_S }{\left(\hat{s}-M_S^{2}\right)^{2}+ M_S^{2}\Gamma_S^{2}
} \,\right) ^{2}  \right] \; .
\label{eu:nud}
\end{eqnarray}
We must also consider the subprocess $e \bar{d} \rightarrow \nu_e \bar{u}$ that
exhibits contributions of  $W$ in the t-channel  and $S$ in the
u-channel, resulting in
\begin{equation}
\left(\frac{d\hat{\sigma}}{d\hat{t}}\right)_{e \bar{d}\rightarrow
\nu_e \bar{u}}  = \frac{1}{64 \pi}
 \frac{\hat{u}^{2}}{\hat{s}^2} 
\left(\,\frac{ g^{2}}{\hat{t}-M^{2}_{W} } + 
\frac{ \lambda^{2} }{\hat{u}-M_S^{2} }\,\right) ^{2} \; .
\label{edbar:nuubar} 
\end{equation}

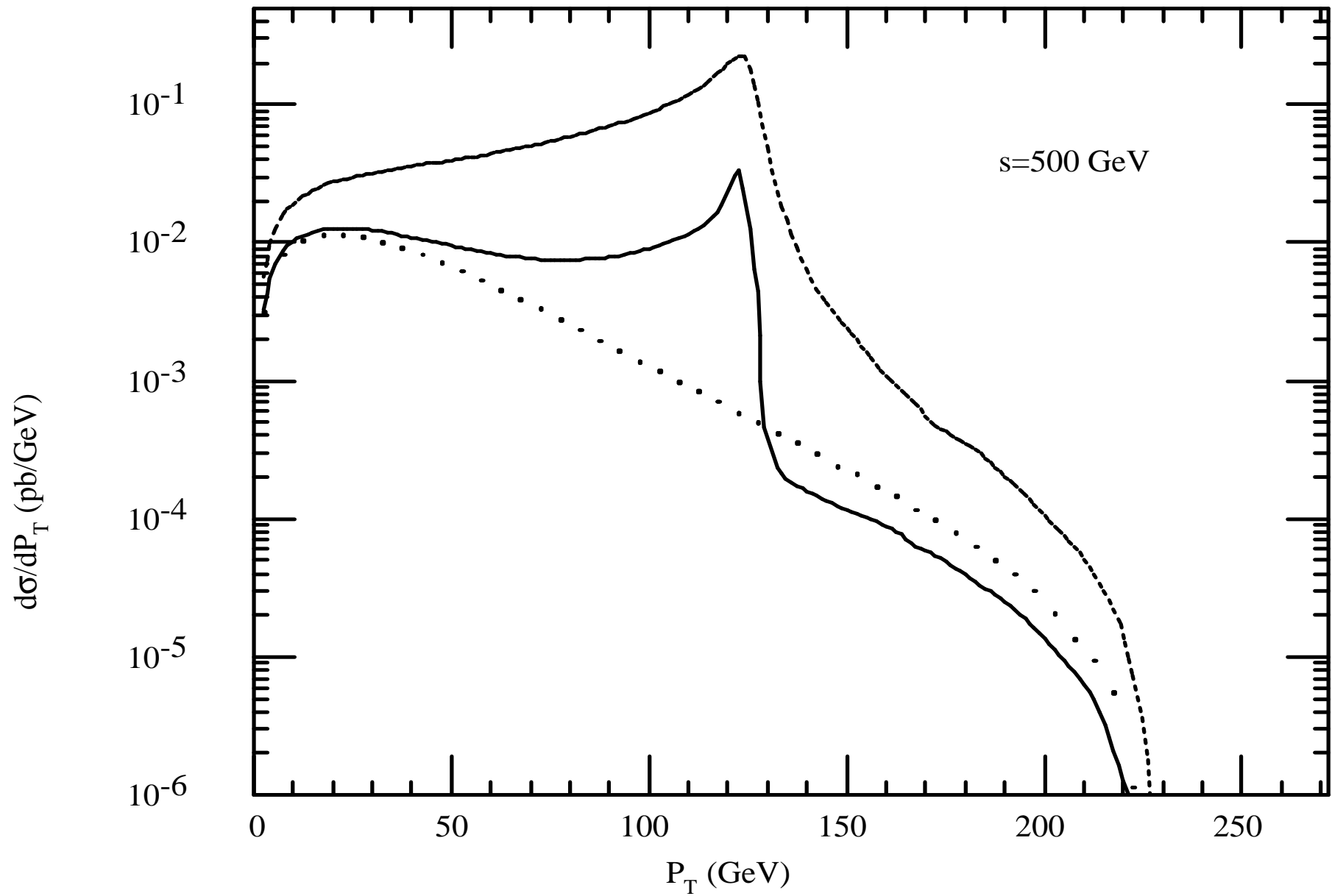
\begin{figure}
\caption{Transverse momentum distribution of the quark (jet)
for the $e^+ e^- \rightarrow e + q_\gamma \rightarrow \mbox{jet} +
\mbox{E}_{\text{missing}}$.  We considered an $e^+ e^- $ collider with
$\protect\sqrt{s} = 500$ GeV and fixed $M_S = 250$ GeV. The curves are for the
leptoquark coupling $\lambda = e$ (solid line), and $\lambda = 1.0$
(dashed line). The dotted line represents the Standard Model
prediction i.e.  $\lambda=0$.}
\label{fig:4}
\end{figure}

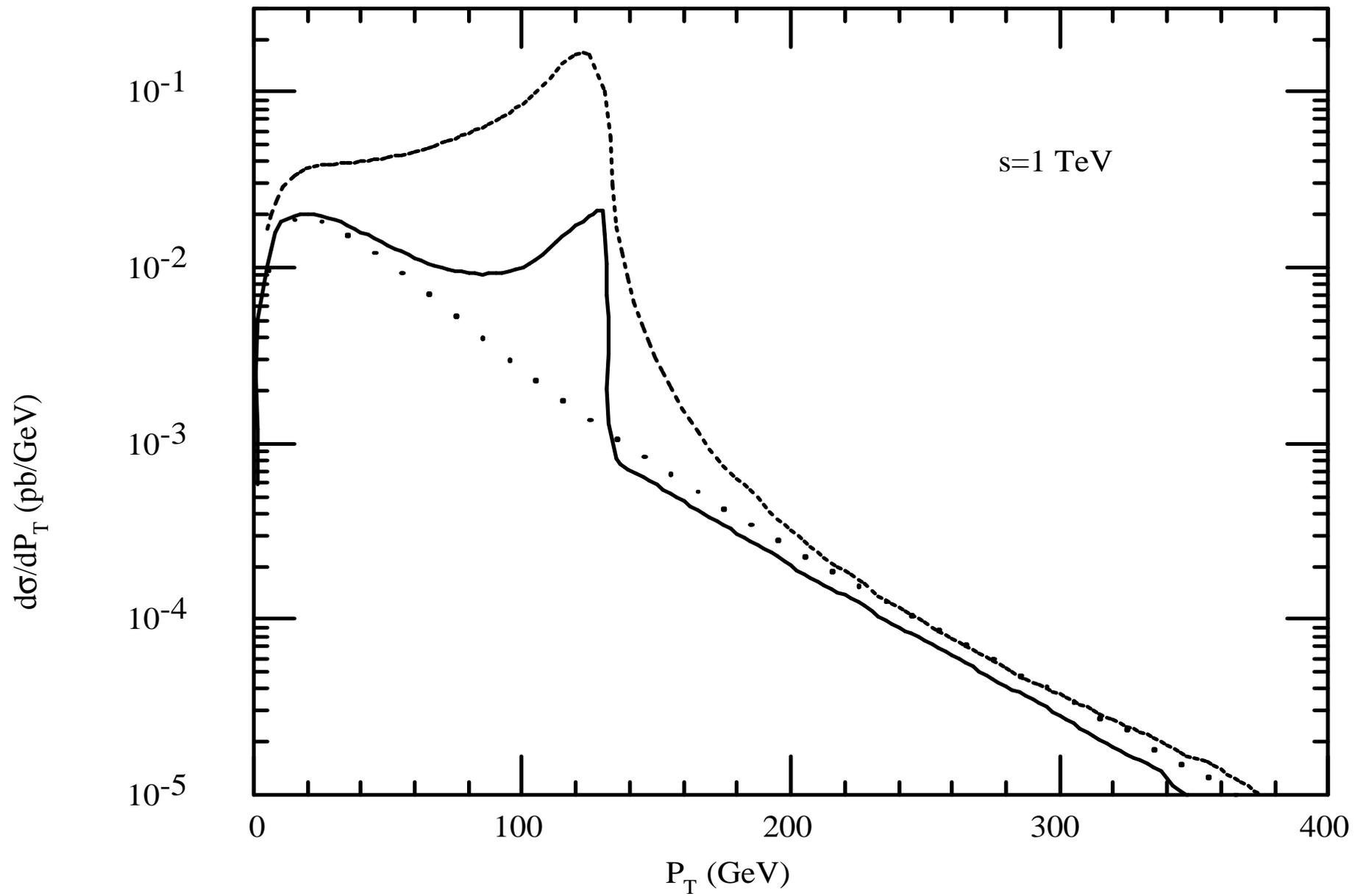
\begin{figure}
\caption{The same as Figure \protect\ref{fig:4} 
for  $\protect\sqrt{s} = 1000$ GeV.} 
\label{fig:5}
\end{figure}

We exhibit in Figs. \ref{fig:4} and \ref{fig:5} the transverse
momentum distribution of the quark (jet) for the process $e\gamma
\rightarrow \text{jet} + \mbox{E}_{\text{missing}}$ for  $\sqrt{s} =
500$ and $1000$ GeV respectively. This distribution is shown for the
Standard  Model (i.e. $\lambda = 0$), $\lambda = e$, and $1$, assuming
$M_S = 250$ GeV. Here again is quite evident the presence of the
leptoquark, and due to the high luminosity expected for the $e\gamma$
machines, we should be able to measure $\lambda$ and $M_S$ with a
reasonable accuracy from the $p_T$ distribution.

For the signal (\ref{s:jet}), we should consider the subprocesses 
$e \gamma \rightarrow S \bar{u}$ and $e g \rightarrow S \bar{u}$.
In the former case there are contributions of the electron
in the s-channel, $u$-quark in the t-channel, and of the $S$ in
the u-channel, leading to
\begin{eqnarray}
\left(\frac{d\hat{\sigma}}{d\hat{t}}\right)_{e \gamma \rightarrow
S\bar{u}}  =  \frac{\lambda^{2}\alpha}{72}
\frac{1}{(-\hat{s}\hat{t})}  \left\{\left[ 2 +
\frac{\hat{u}}{\hat{u}-M_{S}^{2}}  +
\frac{3\left(\hat{u}+\hat{t}\right)}{\hat{s}} \right]^{2} \right.
\left. + \left(\frac{\hat{u}}{\hat{u}-M_S^2} + 
\frac{3\hat{u}}{\hat{s}} \right)^2 \right\} \; .
\label{e:gamma}
\end{eqnarray}
On the other hand, the cross section for $e g \rightarrow S \bar{u}$ receives 
contribution from the $u$-quark in the t-channel and from $S$ in
the u-channel, giving rise to
\begin{eqnarray}
\left(\frac{d\hat{\sigma}}{d\hat{t}}\right)_{e g \rightarrow S \bar{u}} 
= \frac{\lambda^2\alpha_s}{16}  \frac{1}{(-\hat{s}\hat{t})} 
\left[ \frac{\hat{u}^2 + M_S^4}{(\hat{u}-M_S^2)^2} \right] \; .
\label{e:gluon}
\end{eqnarray}

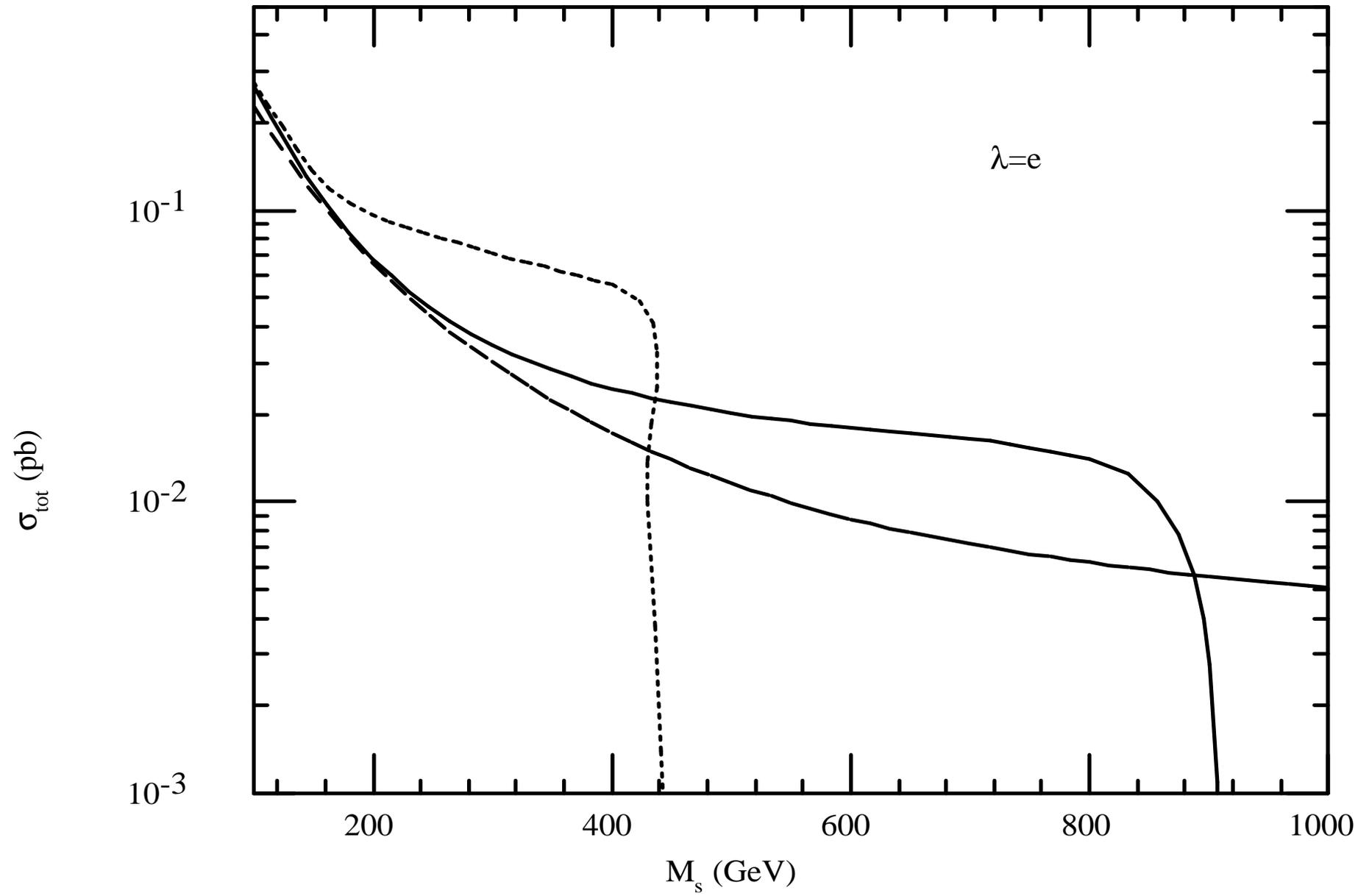
\begin{figure}
\caption{ Total cross section for the process $e^+ e^- 
\rightarrow e + \gamma(g_\gamma)
\rightarrow S + \mbox{jet}$ as a function of $M_S$ with
$\lambda =e$ and $\protect\sqrt{s} = 500$ GeV (dotted line), 
$\protect\sqrt{s} = 1000$ GeV (solid line), and $\protect\sqrt{s} =
2000$ GeV (dashed line). We added the contributions from the
subprocesses (\protect\ref{e:gamma}) and (\protect\ref{e:gluon})}
\label{fig:7}
\end{figure}

Figure \ref{fig:7} exhibits the total cross section, as a function of $M_S$, of
the process $e\gamma \rightarrow S + \text{jet}$ taking into account the
subprocesses (\ref{e:gamma}) and (\ref{e:gluon}), for $\sqrt{s} = 500$, $1000$,
and $2000$ GeV.  In order to regulate the co-linear divergences associated to
the $t$-channel, we required that the angle ($\theta$) of the produced
jet, or of the leptoquark, with the beam pipe satisfies $| \cos \theta | < 0.9$. It is
interesting to notice that this cross section is dominated, for small $M_S$, by
the gluon initiated process, while the direct photon
contribution dominates for large $M_S$.

Once the leptoquark couples to $e u$ and $\nu d$ pairs with the same
strength, the signal for its production in association with a jet is
either $2 ~ \mbox{jets} + e$ or $2 ~ \mbox{jets} +
\mbox{E}_{\text{missing}}$. The main backgrounds for these signals come
from the reactions $e \gamma \rightarrow Z e$, and  $e \gamma
\rightarrow W \nu$ with the $W$ and $Z$ decaying into two jets
\cite{eran}. We can overcome this  background by requiring that the 
invariant mass of the jet pair is not close to $M_{W,Z}$. Another
potential background is the Bethe-Heitler production of hadrons
($\gamma e \rightarrow q \bar{q} e$), which possesses a large cross
section. However, the main contribution to this processes comes from
particles produced with small momenta.  This allows us to reject with
high efficiency this class of events by demanding that the observed
particles and jets have a sufficiently high $p_T$.

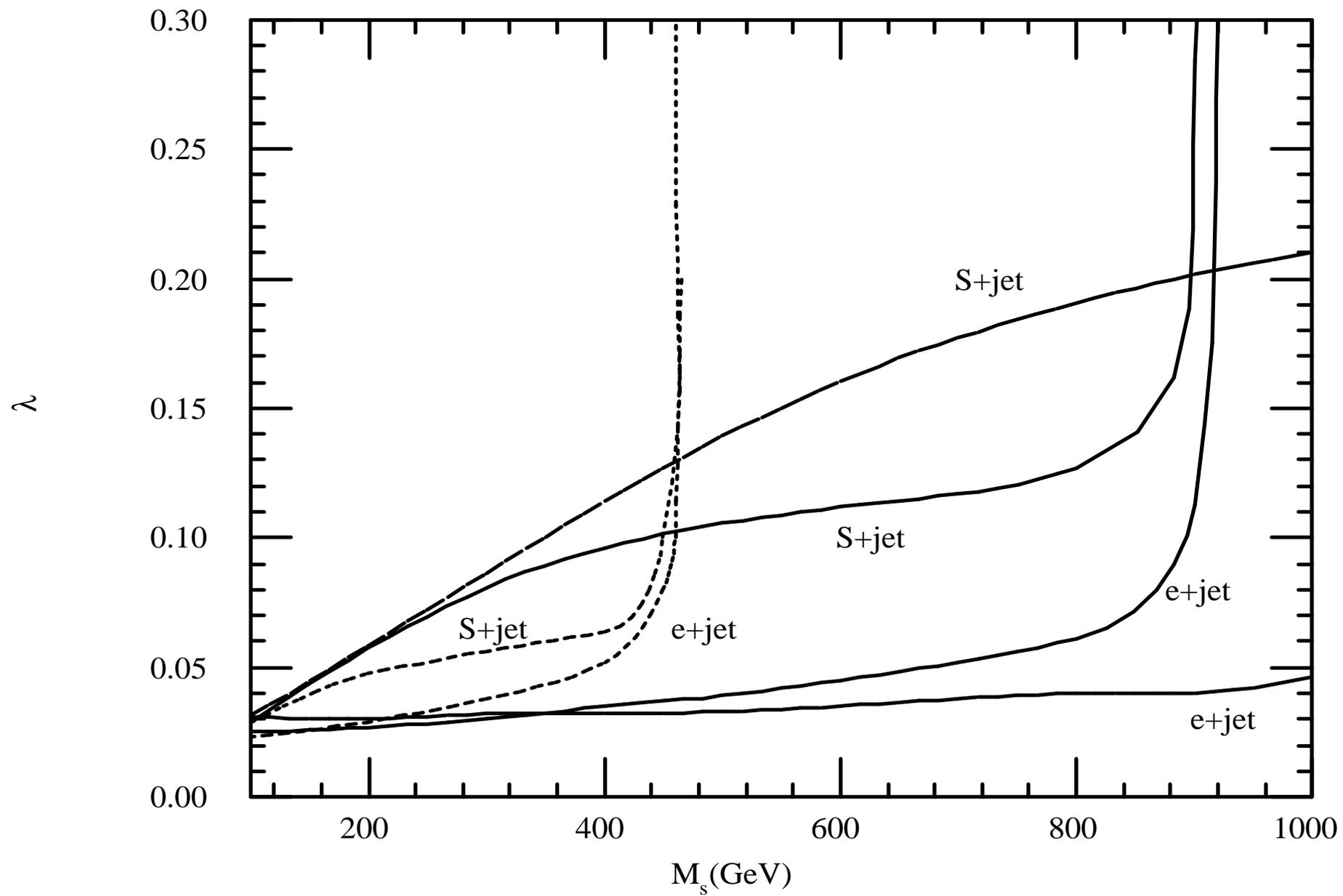
\begin{figure}
\caption{Discovery contour in the plane ($\lambda \times M_S$): (a)  for
the resonant production $e + q_\gamma  \rightarrow  S \rightarrow  e +
\text{jet}$ (lower cuves); (b) for single leptoquark production  $e^+
e^-  \rightarrow e + \gamma(g_\gamma) \rightarrow S + \mbox{jet}$ (upper
curves).  The curves represent an  $e^+ e^- $ collider with
$\protect\sqrt{s} = 500$ GeV (dotted line),  $\protect\sqrt{s} = 1000$
GeV (solid line), and $\protect\sqrt{s} = 2000$ GeV (dashed line). We
assumed a luminosity of $10^4$ pb$^{-1}$ per year.} 
\label{fig:8}
\end{figure}

In order to estimate the capability of the planned $e\gamma$
colliders to search for leptoquarks, we present in Fig.
\ref{fig:8} the discovery contour in the ($\lambda \times
M_S$) plane. For the process (\ref{e:jet}), we required the
significance level of the signal to be $5\sigma$, when we
consider a bin of $\pm 10$ GeV around $M_S$. We can see from
Fig. \ref{fig:8} that we can observe leptoquarks with masses
up to the kinematical limit of the machine provided that the
coupling $\lambda$ is larger than $\sim 5 \times 10^{-2}$.

In Fig. \ref{fig:8} we also present the discovery contour  for single
leptoquark production (\ref{s:jet}), assuming a luminosity of $10^{4}$
pb$^{-1}$ per year for the $e\gamma$ collider.  We required in this case
the occurence of $25$ events per year. The maximum value of leptoquark
mass that can be reached is $M_S \simeq 850$ GeV for a center-of-mass
energy of the $e^+e^-$ machine of $1000$ GeV, taking $\lambda = 0.15$.
Therefore, even for small leptoquark coupling, an $e\gamma$ collider can
investigate a leptoquark almost up to the kinematical limit.

\acknowledgments
This work was partially supported by Conselho Nacional de
Desenvolvimento Cient\'\i fico e Tecnol\'ogico (CNPq), and
Funda\c{c}\~ao de Amparo \`a Pesquisa do Estado de S\~ao Paulo
(FAPESP).

\begin{references}


\bibitem{af} L.\ Abbott and E.\ Farhi, Phys.\ Lett.\ {\bf 101B} (1981)
69; Nucl.\ Phys.\ {\bf B189} (1981) 547.  

\bibitem{tec} S.\ Dimopoulos, Nucl.\ Phys.\ {\bf B168} (1981) 69 ;
E.\ Farhi and L.\ Susskind, Phys.\ Rev.\ D{\bf 20} (1979) 3404; J.\
Ellis {\em et.\ al.\/}, Nucl.\ Phys.\ {\bf B182} (1981) 529.

\bibitem{gut} See, for instance,  P.\ Langacker, Phys.\ Rep.\ {\bf
72} (1981) 185.

\bibitem{e6} J.\ L.\ Hewett and T.\ G.\ Rizzo, Phys.\ Rep.\ {\bf 183}
(1989) 193.

\bibitem{wud} J.\ Wudka, Phys.\ Lett.\ {\bf 167B} (1986) 337.

\bibitem{lepto:ep} J.\  F.\ Gunion and E.\ Ma, Phys.\ Lett.\
{\bf 195B} (1987) 257; A.\ Dobado, M.\ J.\ Herrero and C.\
Munoz, Phys.\ Lett.\ {\bf 191B} (1987) 449.

\bibitem{buch} W.\ Buchm\"uller, R.\ R\"uckl, and D.\ Wyler, Phys.\
Lett.\ {\bf 191B} (1987) 442.

\bibitem{lepto:ha} O.\ J.\ P.\ \'Eboli and A.\ V.\ Olinto,
Phys.\ Rev.\ D{\bf 38} (1988) 3461; A.\ Dobado,  M.\ J.\
Herrero and C.\ Munoz, Phys.\ Lett.\ {\bf 207B} (1988) 97;
J.\ L.\ Hewett and S.\ Pakvasa, Phys.\ Rev.\ D{\bf 37}
(1988) 3165; M.\ de Montigny and L.\ Marleau, Phys.\ Rev.\
D{\bf 41} (1990) 3523. 

\bibitem{lepto:ee1}J.\ L.\ Hewett and T.\ G.\ Rizzo, Phys.\
Rev.\ D{\bf 36}, (1987) 3367; O.\ J.\ P.\ \'Eboli and J.\
E.\ Cieza-Montalvo, Phys.\ Rev.\ D{\bf 47}  (1993) 837; J.\
Blumlein and R.\ Ruckl, Phys.\ Lett.\ {\bf 304B} (1993) 337.

\bibitem{lepto:ee2}J.\ L.\ Hewett and S.\ Pakvasa, Phys.\
Lett.\ {\bf 227B} (1989) 178; H.\ Nadeau and D.\ London,
Univ.\ Montreal Report, UdeM-LPN-TH-132 (1993).

\bibitem{las0} F.\ R.\ Arutyunian and V.\ A.\ Tumanian, Phys.\
Lett.\ {\bf 4} (1963) 176; 
R.\ H.\ Milburn, Phys.\ Rev.\ Lett.\ {\bf 10} (1963) 75.

\bibitem{laser} For a review see, V.\ Telnov Nucl.\ Instr.\ Meth.\
{\bf A294} (1990) 72.

\bibitem{buch:2} W.\ Buchm\"uller and D.\ Wyler, Phys.\ Lett.\ {\bf 177B}
(1986) 377.

\bibitem{cdf} M.\ S.\ Gold (CDF Collab.), talk given at the XXVI
International Conference in High Energy Physics, Dallas (1992).

\bibitem{witten} E.\ Witten, Nucl.\ Phys.\ {\bf B120} (1977) 189.

\bibitem{do} D.\ W.\ Duke and J.\ F.\ Owens, Phys.\ Rev.\ {\bf D26}
(1982) 1600. 

\bibitem{dg} M.\ Drees and K.\ Grassie, Z Phys.\ {\bf C28} (1985) 451.

\bibitem{lac} H.\ Abramowicz, K.\ Charchula, and A.\ Levy, Phys.\
Lett.\ {\bf B269}  (1991) 458.

\bibitem{cor} This cross section has been evaluated in Ref.\cite{wud}.
We reproduce it here, fixing a minor misprint.

\bibitem{eran} The background associated to  bremsstrahlung photons
has been analyzed by  E.\ Yehudai, Phys.\ Rev.\ D{\bf 42} (1990) 771.


\end{references}
\end{document}